\documentclass{ws-spin}
\usepackage{graphicx,psfrag}
\graphicspath{{/home/user/fig/publi/muon/}}
\DeclareGraphicsExtensions{.ps, .eps}
\usepackage{multicol}

\def\gtrsim{\mathrel{\hbox{\lower1ex\hbox{\rlap{$\sim$}\raise1ex\hbox{$>$}}}}}

\begin{document}

\catchline{1}{1}{2015}{}{}
\markboth{Dalmas de R\'eotier {\it et al.}}{Magnetic order, spin waves and fluctuations in the triangular antiferromagnet La$_2$Ca$_2$MnO$_7$}

\title{Magnetic order, spin waves and fluctuations in the triangular antiferromagnet La$_2$Ca$_2$MnO$_7$} 

\author{P. Dalmas de R\'eotier, C. Marin, A. Yaouanc, T. Douce, A. Sikora}
\address{Univ. Grenoble Alpes, INAC/SPSMS, F-38000 Grenoble, France\\
CEA, INAC/SPSMS, F-38000 Grenoble, France}
\author{A. Amato, C. Baines}
\address{Laboratory for Muon-Spin Spectroscopy, 
Paul Scherrer Institute, 5232 Villigen-PSI, Switzerland}

\maketitle

\begin{history}
\received{19 December 2014}
\end{history}

\begin{abstract}

We report magnetic susceptibility, specific heat and muon spin relaxation ($\mu$SR) experiments on the triangular 
antiferromagnet La$_2$Ca$_2$MnO$_7$ which develops a genuine two-dimensional, three-sublattice $\sqrt{3} \times \sqrt{3}$ 
magnetic order below $T_{\rm N} = 2.8$~K. From the susceptibility and specific heat data an estimate of the exchange
interaction is derived. A value for the spin-wave gap is obtained from the latter data. The analysis of a previously reported 
inelastic neutron scattering study yields values for the exchange and spin-wave gap compatible with the results obtained from macroscopic measurements. An appreciable 
entropy is still missing at $10$~K that may be ascribed to intense short-range correlations. 
The critical paramagnetic fluctuations extend far above $T_{\rm N}$, and can be partly understood in terms of two-dimensional 
spin-wave excitations. While no spontaneous $\mu$SR field is observed below $T_{\rm N}$, persistent spin dynamics is found. 
Short-range correlations are detected in this temperature range. Their relation to a  possible molecular spin substructure 
and the observed exotic spin fluctuations is discussed.

\end{abstract}


\begin{multicols}{2}

\section{Introduction} 
\label{Introduction}

The search for new states of matter forms a frontier research area in many-body physics. Geometrically frustrated magnetic materials are a fertile ground for this search.\cite{Ramirez01,Lee08,Balents10} Thanks to the discovery of the high-$T_c$ superconductors nearly 30 years ago, much of the interest has been devoted to two dimensional (2D) systems. It is known theoretically that a 2D Heisenberg antiferromagnet can be either long-range ordered or remain a magnetic liquid at 0~K, depending on the geometry.  A N\'eel order of spins $1/2$ is stable on a square lattice.\cite{Huse88} A non-collinear $\sqrt{3} \times \sqrt{3}$ state, with spins oriented at $120^\circ$ to each other, is believed to be realized for an equilateral triangular lattice,\cite{Huse88,Bernu92,Capriotti99} although this has recently been questioned for spins $1/2$.\cite{Suzuki14} For a kagome lattice of spins $1/2$ there is evidence for a spin-liquid state.\cite{Yan11}

Obviously, no experiment is performed at 0~K. Hence, it is important to recall that no long-range magnetic order is possible above zero temperature for 2D systems with continuous symmetry.\cite{Mermin66a,Mermin66b} However, a symmetry breaking term may lead to magnetic order at finite temperature.\cite{Diep04}

Since the physics of the simple square lattice Heisenberg model is understood, the attention has partly turned to the equilateral triangular lattice Heisenberg system. It has the advantage to be geometrically simpler than the kagome. The interest is in the study of compounds which do not display long-range magnetic order induced by inter-layer magnetic coupling. Only a few candidates have been identified. 

An organic material has attracted much attention, but the triangular lattice is observed to be distorted.\cite{Shimizu03,Pratt11} NiGa$_2$S$_4$ is an insulating antiferromagnetic insulator for which the Ni$^{2+}$ ions are on an exact triangular lattice.\cite{Nakatsuji05} Its ground state is spin disordered. However, the third nearest neighbour interaction dominates according to magnetic neutron diffraction. The sibling compound FeGa$_2$S$_4$ also displays a spin-disordered ground state.\cite{Dalmas12} The range of the exchange interaction has not be determined experimentally.

The Mott insulator NaCrO$_2$ is quite interesting because it crystallizes in the space group $R {\bar 3}{\rm m}$.\cite{Motida70} This means that the triangular spin layers are rhombohedrally stacked. Therefore inter-layer mean fields cancel which leads to a decoupling of the spin layers at the mean-field level. The system is characterized by two temperatures: $T
_1 \simeq 40$~K and $T_2 \simeq 30$~K.\cite{Olariu06} Muon spin relaxation ($\mu$SR) reveals an onset of gradual spin freezing at $T_1$. Nuclear magnetic resonance and $\mu$SR resolve spontaneous fields below $T_2$ corresponding to short-range correlations. A recent neutron study shows a small magnetic incommensuration with four magnetically coupled layers to appear below $T_2$.\cite{Hsieh14}  The in-plane correlation length is limited to $\approx 6$~nm in the same temperature range. If there was no incommensuration, NaCrO$_2$ would be a realization of a magnetic triangular lattice as studied theoretically.\cite{Huse88,Bernu92,Capriotti99}

In this context, the triangular antiferromagnet La$_2$Ca$_2$MnO$_7$ is of interest.\cite{Bao09} As NaCrO$_2$, it crystallizes in the space group $R {\bar 3}{\rm m}$,\cite{Wang04} with lattice parameters $a =5.619$~\AA\ and $c =17.292$~\AA\ at 0.04~K. The half-filled $t^3_{\rm 2g}$ electronic configuration of Mn$^{4+}$ has a spin $S=3/2$. In Wyckoff notations, this ion is at position 3a (hexagonal setting) of point symmetry ${\bar 3}m$, with the three-fold axis perpendicular to the triangular plane.

A wealth of data is already available.\cite{Bao09} Susceptibility and specific heat data indicate that a second-order magnetic phase transition occurs at $T_{\rm N} = 2.8$~K. According to elastic powder neutron diffraction, the $\sqrt{3} \times \sqrt{3}$ spin order is realized below $T_{\rm N}$. It is 2D-like, as certified by the Warren diffraction profiles, in  defiance of the Mermin-Wagner theorem  which states that no ordering should occur at finite temperature for a system with continuous spin symmetry.\cite{Mermin66a,Mermin66b} Inelastic neutron powder scattering shows the magnetic order to support spin-waves. 

Here we present and analyze magnetic susceptibility and specific heat data of La$_2$Ca$_2$MnO$_7$. They give access to the exchange interaction which can be compared to the value extracted from the spin-wave dispersion measured by neutron scattering. A spin gap is derived from specific heat and neutron scattering data. The spin dynamics is studied with $\mu$SR spectroscopy. The critical paramagnetic spin-lattice relaxation is analyzed in terms of 2D spin waves. Surprisingly, no spontaneous precession of the muon spin is detected below $T_{\rm N}$. This is taken as a signature of the dynamical nature of the ground state. Short-range magnetic correlations are inferred from the analysis of spectra in the ordered state.

The organization of this paper is as follows. Section~\ref{Experimental} introduces the experimental methods. In Secs.~\ref{Bulk} and \ref{Muon} are presented and analyzed experimental results from bulk and $\mu$SR measurements, respectively. In the following section (Sec.~\ref{Discussion}) a discussion of our results is provided. Conclusions are gathered in Sec.~\ref{Conclusions}. 

\section{Experimental} 
\label{Experimental}

The experimental work has been done with powder samples. They have been  prepared through a chemical nitrates route using as starting materials: La(NO$_3$)$_3\,\cdot\,6$H$_2$O (4N), Ca(NO$_3$)$_2 \,\cdot\, \frac{5}{2}$H$_2$O (4N7), and Mn(NO$_3$)$_2 \,\cdot\,$5H$_2$O (5N). The calcination step has been held at 650$^\circ$C for 1~hour before several grindings and heat treatments up to 900$^\circ$C under air. Additional heat treatments at 900$^\circ$C under argon and then oxygen atmosphere have also been performed on small pellets. The phase purity has been checked by X-ray diffraction and no foreign phases have been detected. 

Magnetic susceptibility measurements have been performed with a commercial (Quantum Design MPMS) magnetometer. Specific heat data have been recorded with a commercial calorimeter (Quantum Design PPMS) equipped with a $^3$He stage and using a standard thermal relaxation method. $\mu$SR measurements have been carried out at the Low Temperature Facility and General Purpose Surface muon spectrometer of the Swiss Muon Source (Paul Scherrer Institute, Switzerland).

\section{Results from macroscopic measurements} 
\label{Bulk}

In Fig.~\ref{La2Ca2MnO7_suscep} is displayed the magnetic susceptibility versus temperature, i.e.\ $\chi (T)$, at low temperature. It is measured under an external field $B_{\rm ext} = 0.5$~T. A maximum is observed at $T_{\rm N} = 2.88 \, (5)$~K, in agreement with previous data,\cite{Bao09} but here the maximum is more pronounced, and $\chi (T)$ is slightly smaller although measured with the same $B_{\rm ext}$.

\begin{figurehere}
\begin{center}
\includegraphics[scale=0.70]{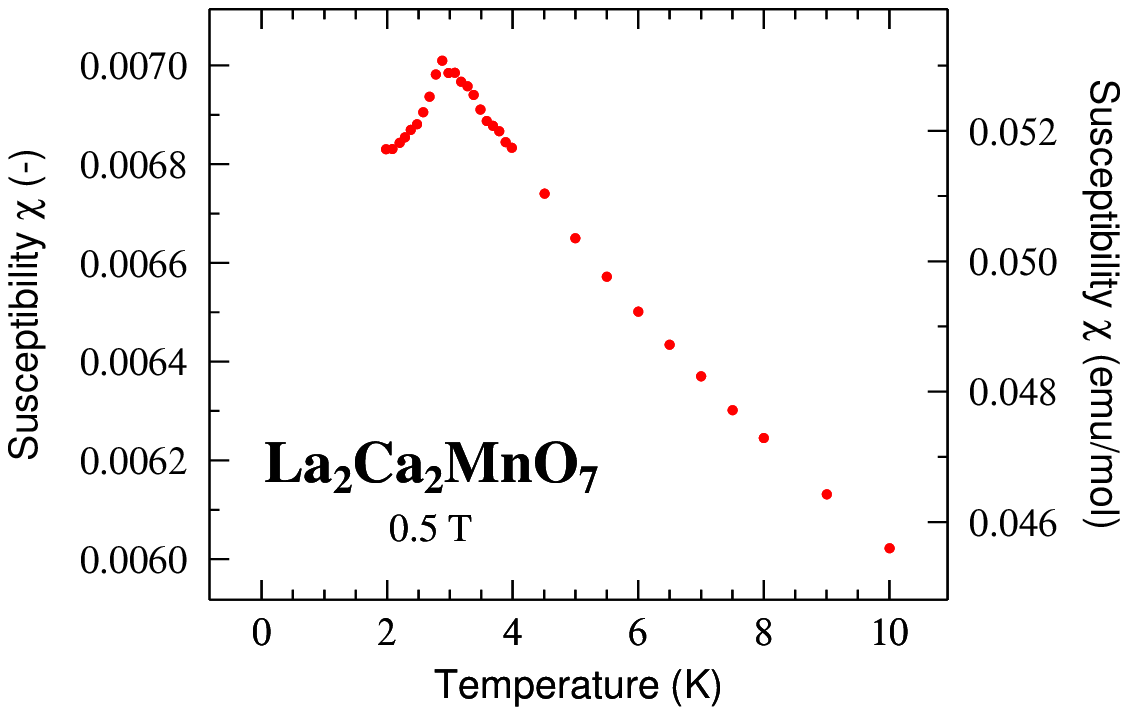}
\end{center}
\caption{
Thermal dependence of the magnetic susceptibility of our powder of La$_2$Ca$_2$MnO$_7$ at $B_{\rm ext} = 0.5$~T. The data are presented in SI and electro-magnetic units for ease of comparison.}
\label{La2Ca2MnO7_suscep}
\end{figurehere}

As shown in  Fig.~\ref{La2Ca2MnO7_suscep_inv}, $1/\chi (T)$ exhibits a Curie-Weiss behaviour for $T < 25$~K with a Curie-Weiss temperature $\theta_{\rm CW} = -41 \ (1)$~K, at variance with $\theta_{\rm CW} = -25$~K given previously.\cite{Bao09} We compute for the empirical Ramirez parameter of frustration $f_{\rm R} = |\theta_{\rm CW} |/T_{\rm N} = 14 \gg 1$, indicating strong frustration.\cite{Ramirez01} Considering the Heisenberg Hamiltonian ${\mathcal H}_{\rm H} = J \sum_{\langle i,j, \rangle} {\bf S}_i \cdot {\bf S}_i$, the exchange constant $J$ of a spin with its $z =6$ neighbours can be evaluated in the mean-field approximation:
\begin{equation}
J = J_\theta = {3 k_{\rm B} |\theta_{\rm CW}| \over 2 z S(S +1)}.
\label{Curie_Weiss}
\end{equation}
From the $\theta_{\rm CW}$ value, we compute $J_\theta = 0.24 \, (1)$~meV. 
\begin{figurehere}
\begin{center}
\includegraphics[scale=0.70]{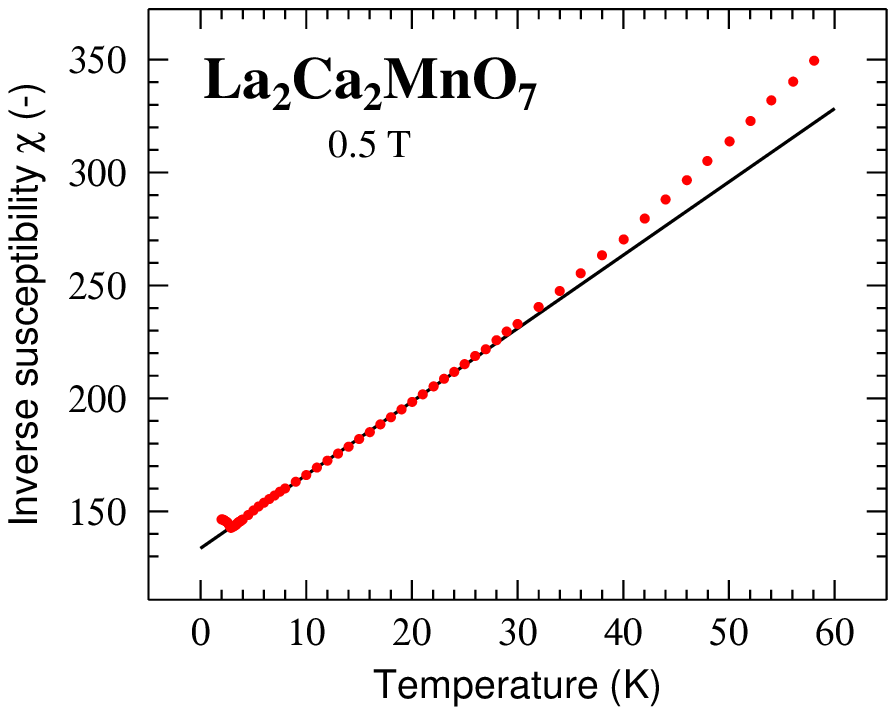}
\end{center}
\caption{
Inverse of the susceptibility of La$_2$Ca$_2$MnO$_7$ versus temperature at $B_{\rm ext} = 0.5$~T. The solid line results from a fit with a Curie-Weiss law for $T < 25$~K as explained in the main text. Here we use SI units.
}
\label{La2Ca2MnO7_suscep_inv}
\end{figurehere}

We present in Fig.~\ref{La2Ca2MnO7_heat_global} the measured specific heat $C_{\rm p}$ at low temperature. It displays a maximum at $2.81 \, (3)$~K, in agreement with the $T_{\rm N}$ value deduced from $\chi(T)$. In the previously reported data,\cite{Bao09} $C_{\rm p}(T_{\rm N})$ is $\approx 22 \, \% $ larger and  decreases markedly for $T > T_{\rm N}$ up to $8.0$~K, the high-temperature limit of the measurements. Our data are clearly different. The entropy variation plotted in the insert of Fig.~\ref{La2Ca2MnO7_heat_global} shows that only $\approx 70 \, \%$ of the available entropy, i.e.\ $R \ln (4)$, is recovered at 10~K. Because the compound orders magnetically, the entropy is zero at $T=0$. Since the lattice should contribute above $ \approx 6$~K, we deduce that more than $30 \, \%$ of the entropy is still missing at 10~K. While for a definite conclusion about this missing fraction, measurements on La$_2$Ca$_2$MnO$_7$ above 10~K and on a non-magnetic isostructural compound need to be performed, this result suggests the existence of dynamic short-range spin correlations persisting far above $T_{\rm N}$. They could be of chiral nature.

The phonon contribution to $C_{\rm p}$ is fully negligible below $T_{\rm N}$. The measurements have not been performed at sufficiently low temperature for any nuclear contribution to be of importance. This means that $C_p$ should only be of spin-wave origin for the data of Fig.~\ref{La2Ca2MnO7_heat_global} at $T < T_{\rm N}$. As successfully  done recently,\cite{Dalmas12a} phenomenologically we write for the dispersion relation of the spin waves 
\begin{figurehere}
\begin{center}
\includegraphics[scale=0.80]{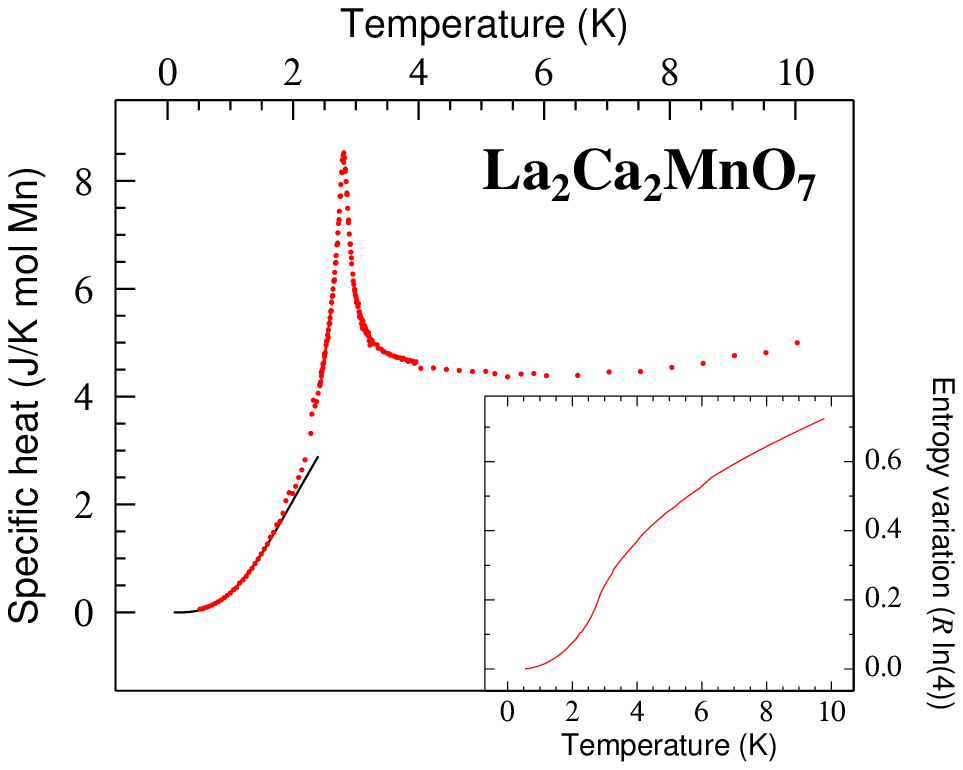}
\end{center}
\caption{
Thermal dependence of the specific heat $C_{\rm p}$  of La$_2$Ca$_2$MnO$_7$ from $0.52$~K upwards. The solid line represents a fit of the low temperature data  as explained in the main text. The fit is excellent for $T \leq 1.6$~K. In the insert is pictured the entropy variation from $0.54$ to 10~K normalized to $R \ln (4)$. 
}
\label{La2Ca2MnO7_heat_global}
\end{figurehere}
\begin{equation}
\hbar^2 \omega^2({\bf q}) = \Delta^2_{\rm gap} + \hbar^2 {\tilde \omega^2}({\bf q}), 
\label{gap}
\end{equation}
where $\Delta_{\rm gap}$ is the spin gap and  $\hbar {\tilde \omega}({\bf q})$ the spin-wave energy without the gap which only depends on the exchange energy.\cite{Jolicoeur89} The fit pictured in Fig.~\ref{La2Ca2MnO7_heat_global} gives  $\Delta_{\rm gap} = \Delta_{\rm sh} = 0.17 \, (1)$~meV and $J = J_{\rm sh} = 0.22 \, (1)$~meV. 

\section{$\mu$SR results} 
\label{Muon}

We have performed $\mu$SR measurements from above $T_{\rm N}$ down to 0.04~K. Basically such an experiment\cite{Dalmas97,Dalmas04,Yaouanc11} consists in the implantation of polarized muons in the sample under study. Muons are unstable spin 1/2 particles with a lifetime $\tau_\mu$ = 2.2~$\mu$s. Through the detection of the decay positrons one measures the so-called polarization function $P_Z^{\rm exp}(t)$ which monitors the evolution of the projection of the muon spin ensemble along the direction $Z$ of the initial polarization. In the magnetically ordered state we should observe in zero field (ZF) for our powder sample the weighted sum of a damped oscillation corresponding to the precession of the muons in the spontaneous field at the muon site and a simple relaxation for the response of the muons for which the local field is parallel to their initial polarization. In the paramagnetic state an exponential relaxation characterized by the spin-lattice relaxation rate $\lambda_Z$ is expected. 

On general ground, the experimental signal, denoted as asymmetry, is the sum of two components, i.e.\ $a_0 P^{\rm exp}_Z(t) = a_{\rm s} P_Z(t) + a_{\rm bg}$, where the first component accounts for the response of the sample and the second is a time-independent background, reflecting the muons stopped in the sample vicinity.

We shall first expose our results obtained below  $T_{\rm N}$. Then we shall consider the paramagnetic state.

\subsection{Ordered phase} 
\label{Muon_Ordered}

Two spectra recorded for $T < T_{\rm N}$ are pictured in Fig.~\ref{La2Ca2MnO7_muon_spectra_ZF}. Surprisingly, no signature of a spontaneous field is detected, i.e.\ there is neither a spontaneous oscillation nor a drop in the signal amplitude. Therefore the local field at the muon site(s) does not keep a constant value a time sufficiently long for spontaneous oscillations to be observed. Although the model in Ref.~\refcite{Dalmas06} predicts an exponential relaxation in the fast spin dynamics limit, in contrast to observation, qualitatively $\nu_{\rm s} > \gamma_\mu B_{\rm fluct}$. Here $\nu_{\rm s}$ is the fluctuation frequency of the spontaneous fluctuating field ${\bf B}_{\rm fluct}$ and $\gamma_\mu$ = 851.615~Mrad\,s$^{-1}$\,T$^{-1}$ is the muon gyromagnetic ratio. Since $B_{\rm fluct}$ is in the range $0.1-0.2$~T,\cite{Dalmas12a} we compute $\nu_{\rm s} \gtrsim 10^8$~s$^{-1}$. 
\begin{figurehere}
\begin{center}
\includegraphics[scale=0.80]{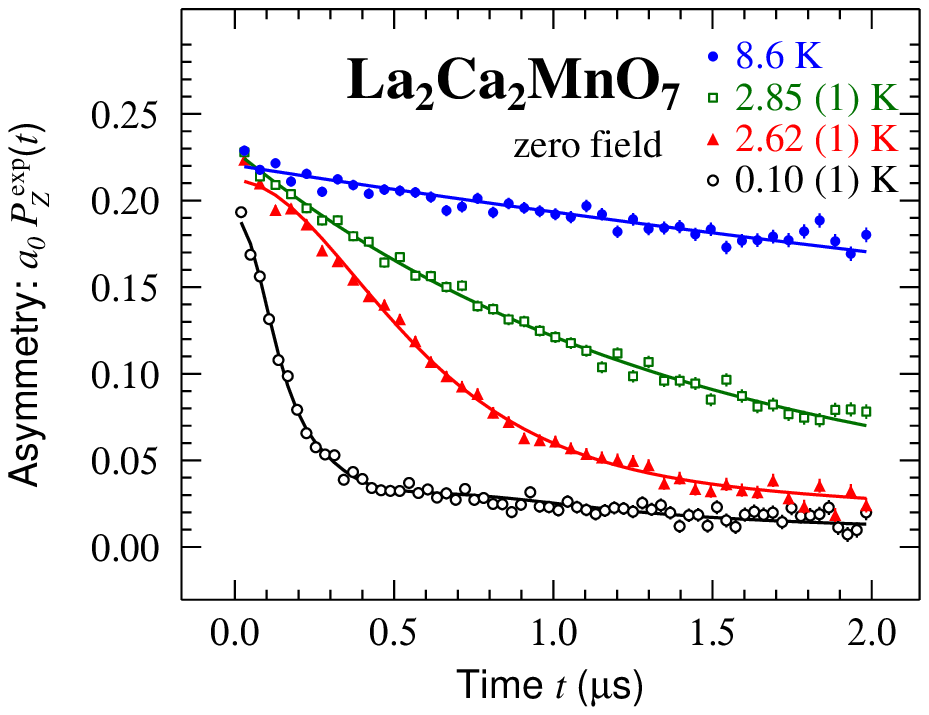}
\end{center}
\caption{
A selection of zero-field $\mu$SR spectra for a powder of La$_2$Ca$_2$MnO$_7$. The two low temperature spectra were recorded in the magnetically ordered state. Solid lines depict fits described in the main text. Note that the $0.10$~K spectrum on one hand and the other three spectra on the other hand have been recorded with different spectrometers: this explains the slight difference in the signal amplitudes. }
\label{La2Ca2MnO7_muon_spectra_ZF}
\end{figurehere}

The asymmetry spectra are described by the polarization function $P_Z(t) = P_Z(\delta_{\rm ZF},\nu_{\rm c},\eta_3,\eta_4,t)$, where
$\delta_{\rm ZF}$ is the width of the field distribution at the muon site and $\nu_{\rm c}$ stands for the correlation frequency of the magnetic fluctuations. The two parameters $\eta_3$  and $\eta_4$ characterize short-range magnetic correlations, if any. We are in fact using the so-called dynamical Kubo-Toyabe model,\cite{Hayano79} recently extended to account for short-range correlations.\cite{Yaouanc13a} The parameter $\delta_{\rm ZF}$ is the field standard deviation $\Delta_{\rm ZF}$ when $\eta_3 = \eta_4 = 0$. Here a proper description of the spectra (Fig.~\ref{La2Ca2MnO7_muon_spectra_ZF}) is obtained with $\eta_3 = 0.74 \, (2)$ and $\eta_4 = 0.47 \, (2)$, common to all the spectra. In Fig.~\ref{La2Ca2MnO7_muon_parameters_ZF} we have plotted $\delta_{\rm ZF} (T)$ and $\nu_{\rm c} (T)$. The thermal behaviour of $\delta_{\rm ZF}$ is conventional for a second order phase transition, although its size at low temperature is anomalously small to reflect the magnetic moments of Mn$^{4+}$. $\nu_{\rm c} (T)$ is exotic since the spin dynamics is not frozen at low temperature. This is the signature of persistent spin dynamics, a characteristics of geometrically frustrated materials observed for compounds with\cite{Bertin02,Zheng05,Lago05,Yaouanc05a,Dalmas06,Chapuis09b,Yaouanc13} and without\cite{Uemura94,Gardner99,Keren00,Hodges02,Lago05,Marcipar09,Yaouanc08,Dalmas12} long-range magnetic order. An other description of the ZF spectra is possible. It is mathematically different, but physically similar, although the existence of short-range correlations cannot be inferred from it.\footnote{It is inspired by nuclear magnetic resonance practice, i.e.\ $P_Z(t) = {1 \over 3} \exp \left ( -\lambda_Z t \right ) + {2 \over 3} \exp \left (- {\gamma_\mu^2 \Delta_X^2  t^2 \over 2} \right )$. Here $\lambda_Z$ is the spin-lattice relaxation rate, $\Delta_X$ the standard deviation of the field distribution at the muon site probed by the second component. If the Gaussian term was multiplied by a cosine function accounting for a spontaneous oscillation, we would have a conventional formula.\cite{Yaouanc11} $\Delta_X(T)$ behaves as an order parameter for a second order phase transition, and $\lambda_Z (T)$ increases slightly from $0.8~\mu {\rm s}^{-1} $ near $T_{\rm N}$ to  $1.8~\mu {\rm s}^{-1} $ at $0.04$~K.}
\begin{figurehere}
\begin{center}
\includegraphics[scale=0.75]{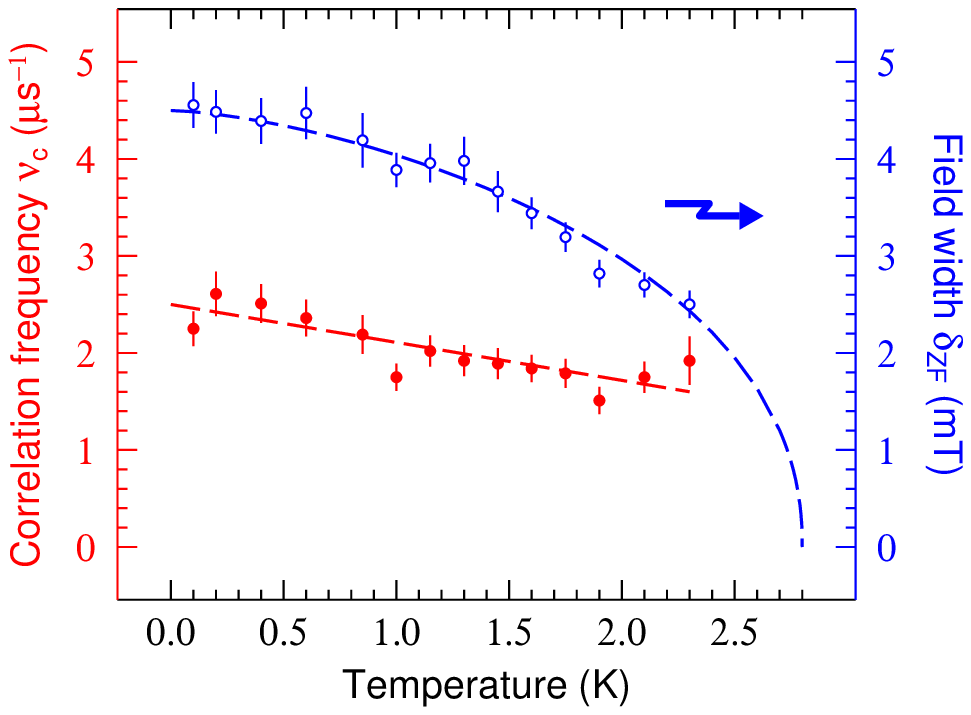}
\end{center}
\caption{
$\delta_{\rm ZF} (T)$ and $\nu_{\rm c} (T)$ in zero field for La$_2$Ca$_2$MnO$_7$. These data refer to the temperature range $T < T_{\rm N}$.
The dashed lines are guides to the eyes.}
\label{La2Ca2MnO7_muon_parameters_ZF}
\end{figurehere}

A physical discussion of these results is given in Sec.~\ref{Discussion_below}.

\subsection{Paramagnetic phase} 
\label{Muon_Para}

Zero and longitudinal field measurements have been performed in the paramagnetic state; see Fig.~\ref{La2Ca2MnO7_muon_spectra_ZF} for two ZF spectra. The spectra are well accounted for by the polarization function $P_Z(t)$ = $\exp(-\lambda_Z t)$. A field scan from $0$ to $400$~mT at $10$~K shows that $\lambda_Z$ is field independent for $B_{\rm ext} \geq 5$~mT. A larger value of $\lambda_Z$ for $B_{\rm ext} < 5$~mT suggests an appreciable contribution of the $^{55}$Mn and $^{139}$La nuclei to the relaxation which is already quenched under $B_{\rm ext} = 5$~mT. This is an expected behaviour.\cite{Yaouanc11} A systematic temperature scan has therefore been carried out for $B_{\rm ext} = 10$~mT. The semi-log plot of $\lambda_Z/T^3$ versus $1/T$ follows a straight line for $T < 11$~K as shown in Fig.~\ref{La2Ca2MnO7_muon_lambdaZ}. This means that 
\begin{figurehere}
\begin{center}
\includegraphics[scale=0.75]{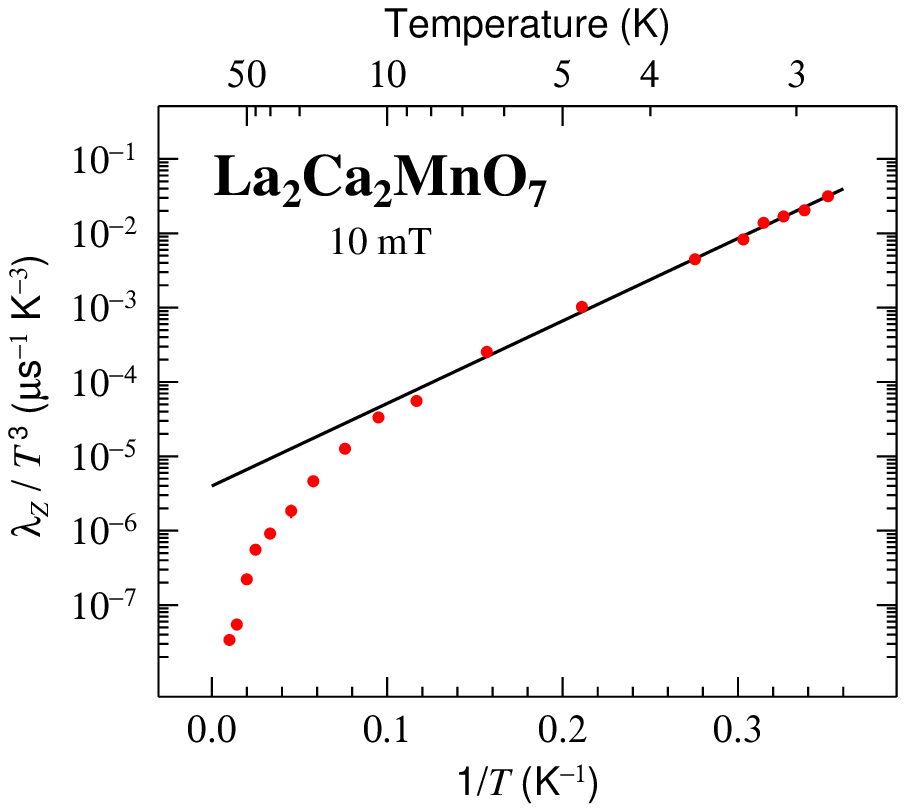}
\end{center}
\caption{Plot of the $\mu$SR spin-lattice relaxation rate $\lambda_Z$ divided by $T^3$ versus $1/T$. The measurements have been performed
in the paramagnetic phase of La$_2$Ca$_2$MnO$_7$ under a longitudinal field $B_{\rm ext} = 10$~mT. The straight line is the prediction of 
a model as detailed in the main text.
}
\label{La2Ca2MnO7_muon_lambdaZ}
\end{figurehere}
\begin{equation}
\lambda_Z \propto T^3 \exp(T_0/T),
\label{fit_2}
\end{equation}
with $T_0$ = 26\,(1)~K. This law is predicted for the $\lambda_Z$ critical behaviour of a 2D frustrated quantum Heisenberg antiferromagnet in the renormalized-classical regime, i.e.\ $T \ll T_0/2$;\cite{Chubukov94,Chubukov94a,Chubukov94b} see also Refs.~\refcite{Azaria92,Lecheminant95}. This boundary condition is fullfilled in our case. As summarized elsewhere,\cite{Dalmas12,Zhao12} $k_{\rm B} T_0 = 4 \pi \rho_{\rm s}$ where $\rho_{\rm s}$ is the spin stiffness constant which can be expressed in terms of $J$ and $S$. We compute $J = J_{\lambda_Z} = 0.16 \, (1)$~meV. This result is further discussed in Sec.~\ref{Discussion_above}.

\section{Discussion} 
\label{Discussion}

\subsection{The energy scales: exchange constant and spin gap} 
\label{Discussion_exchange}

In this work the exchange constant $J$ has been obtained from the paramagnetic susceptibility and the low temperature specific heat, with the results  $J_\theta = 0.24 \, (1)$ and $J_{\rm sh} = 0.22 \, (1)$~meV. A value for the spin gap has been extracted from the latter measurement: $\Delta_{\rm sh} = 0.17 \, (1)$~meV. $J_{\lambda_Z}$ derived from the temperature dependence of the critical paramagnetic fluctuations will be discussed in Sec.~\ref{Discussion_above}. An energy versus wavevector map of the neutron scattering intensity in the vicinity of a point in the Brillouin zone corresponding to a magnetic Bragg reflection has been reported.\cite{Bao09} This will also enable us to get values for the exchange and spin gap energies.

Hypotheses are explicit for our two estimates of $J$. The analysis of the susceptibility is based on the Curie-Weiss formula which is of mean-field nature. Hence, for a triangular lattice expected to display quantum effects, $J_\theta$ may not be reliable. However, it occurs to be consistent with $J_{\rm sh}$. For the analysis of the specific heat we have assumed the linear spin-wave theory to be valid. The renormalization due to spin-wave interactions is predicted to be extremely mild in the vicinity of the three inequivalent points in the Brillouin zone where ${\tilde \omega}({\bf q})$ vanishes.\cite{Dombre89,Chubukov94} Hence, the influence of spin-wave interactions on the determination of $J_{\rm sh}$ should be mild.

A finite spin gap has been derived from the analysis of the low temperature specific heat. The same methodology has been successful in the past to obtain a gap in the same range for Er$_2$Ti$_2$O$_7$.\cite{Dalmas12a} Later on, this estimate has been validated by its determination through inelastic neutron scattering measurements.\cite{Ross14} No theoretical prediction exists for La$_2$Ca$_2$MnO$_7$. 

From the examination of the front locus of the inelastic neutron intensity distribution measured by Bao {\it et al.} at $0.04$~K in the vicinity of the Bragg point $q_0 = 4 \pi/(3 a)$, using Eq.~\ref{gap} with ${\tilde \omega}({\bf q}) = {\tilde \omega}(q) = v_{\rm sw} q$ where $v_{\rm sw}$ is the spin-wave velocity and $q$ is the wavevector measured from $q_0$, we have determined $v_{\rm sw} = v_{\rm n} \approx 600 \, {\rm m/s}$ and $\Delta_{\rm gap} = \Delta_{\rm n} \approx 0.2$~meV. Because these two numbers have been obtained directly from a color map and obviously without taking the resolution function of the spectrometer into account, they can only be taken as a rough estimate. Working with the linear spin-wave approximation and considering that the measurements have been performed in the vicinity of $q_0$,\cite{Dombre89,Chubukov94}
\begin{equation}
J = J_{\rm n} = {2 \over 3} \sqrt{2 \over 3} {\hbar \, v_{\rm n} \over S a} = {4 \over 9} \sqrt{2 \over 3} {\hbar \, v_{\rm n} \over a}, 
\label{spin_wave_velocity_1}
\end{equation}
and therefore $J_{\rm n} \approx 0.25$~meV. 

The three exchange and the two spin gap estimates are rather consistent. Remarkably, the ratio of the gap over the exchange is relatively large, about $0.8$. 

\subsection{Spin dynamics above $T_{\rm N}$} 
\label{Discussion_above}

As expected for a 2D Heisenberg antiferromagnet, the paramagnetic critical spin dynamics as measured by $\lambda_Z$ occurs in a wide temperature range and is well described by Eq.~\ref{fit_2}. The same thermal behaviour has been reported for Li$_2$RuO$_7$ via $^7$Li nuclear magnetic resonance,\cite{Itoh09} and for FeGa$_2$S$_4$,\cite{Dalmas12,Zhao12} NiGa$_2$S$_4$\cite{Dalmas12} and $\alpha$-${\rm KCrO}_2$\cite{Xiao13} by $\mu$SR measurements. However, $J$ extracted from the fit, i.e.\ $J_{\lambda_Z}$, is only approximately $70 \, \%$  of the value derived from the specific heat which is probably the most reliable estimate obtained of $J$ in this study; see Sec.~\ref{Discussion_exchange}. It has been established experimentally that spin excitations exist well above $T_{\rm N}$ at $30$~K,\cite{Bao09} and therefore should contribute to the relaxation. Our analysis is based on a description which neglects $Z_2$ vortex excitations,\cite{Kawamura10,Kawamura11} the density of which may be sufficiently large to influence $\lambda_Z$, leading to a renormalization factor which could explain our results.\cite{Caffarel01,Itoh09} In addition, we would like to note that only the first order correction to the linear spin wave approximation is taken into account.\cite{Chubukov94} Interestingly, while the spin-wave mechanism by itself explains $\lambda_Z(T)$ for FeGa$_2$S$_4$\cite{Dalmas12,Zhao12} and NiGa$_2$S$_4$\cite{Dalmas12}, $J$ extracted from the critical relaxation for Li$_2$RuO$_7$,\cite{Itoh09}  $\alpha$-${\rm KCrO}_2$,\cite{Xiao13} and La$_2$Ca$_2$MnO$_7$ is smaller than expected. Additional theoretical work is necessary. In particular a quantitative estimate of $\lambda_Z$ due to unbound $Z_2$ vortices is required. 

\subsection{Spin dynamics below $T_{\rm N}$} 
\label{Discussion_below}

The analysis presented in Sec.~\ref{Muon_Ordered} assumes the field distribution at the muon site to be isotropic. Since the parameters $\eta_3$ and $\eta_4$ are appreciable, the component field distribution deviates from the Gaussian shape. This means that short-range correlations are present in the ordered state.\cite{Yaouanc13a} Such correlations have first been reported for the pyrochlore antiferromagnet Er$_2$Ti$_2$O$_7$ by neutron diffraction.\cite{Ruff08} Using the same method as here, short-range correlations have been finger-printed in the ordered state of Yb$_2$Ti$_2$O$_7$\cite{Yaouanc13a,Dalmas14} and  Yb$_2$Sn$_2$O$_7$.\cite{Maisuradze15}
 
From the absence of a spontaneous field, we have deduced the presence of fluctuations with fluctuation frequency $\nu_{\rm s} \gtrsim 10^8$~s$^{-1}$. On the other hand, from the analysis of ZF spectra we have inferred a correlation frequency $\nu_{\rm c} \simeq 2.5~\mu{\rm s}^{-1}$ at low temperature.  Hence, it seems there are at least two frequency scales for the magnetic fluctuations with a ratio $\nu_{\rm c}/\nu_{\rm s}$ smaller than $\approx 0.03$. 

The root mean square $\Delta_{\rm ZF}$ of the field at the muon site derived from the measured $\delta_{\rm ZF}$, $\eta_3$ and $\eta_4$ parameters at low temperature is equal to 7.0\,(3)~mT. It is too small to originate from electronic magnetic moments. A possibility is that sporadic spin dynamics is present which would renormalize $\Delta_{\rm ZF}$, as previously reported for the kagome system SrCr$_8$Ga$_4$O$_{19}$, for which the $\mu$SR spectra are also Gaussian-like.\cite{Uemura94}

\subsection{Possible origin of the exotic spin dynamics} 
\label{Exotic}

We have found deep in the paramagnetic state a large missing entropy that suggests intense short-range correlations. Our analysis of the $\mu$SR spectra below $T_{\rm N}$ again indicates the presence of short-range correlations. Hence, there is some type of molecular spin substructures present in La$_2$Ca$_2$MnO$_7$. An obvious structure consists of the three spins on an equilateral triangle. In the ordered state the mean field from these spins is zero. It is tempting to attribute the absence of a spontaneous field to this mean-field feature. Persistent spin dynamics is detected in the ordered state. If it was temperature independent, it could arise from an unidimensional spin structure.\cite{Yaouanc05a,Yaouanc14} However, it is temperature dependent. It would be of interest to determine whether it could be explained by the three spins substructure. The existence of a molecular spin substructure and anomalous slow and persistent fluctuations are obviously closely related.\cite{Yaouanc14,Dally14}

\subsection{Origin of the phase transition} 
\label{Order}

Susceptibility and specific heat indicate a magnetic transition at a non-zero $T_{\rm N}$ value. Neutron diffraction shows the transition to be to a long-range 2D $\sqrt{3} \times \sqrt{3}$ order. Although spontaneous oscillations are not detected below $T_{\rm N}$, the shape of the asymmetry $a_0P^{\rm exp}_Z(t)$ provides a clear signature of the transition as seen in Fig.~\ref{La2Ca2MnO7_muon_spectra_ZF}. Whereas at $T =2.62 \, (1) $~K the curvature of the asymmetry is downwards at short times, it is upwards for $T = 2.85 \, (1)$~K. Hence, the transition takes place between these two temperatures. It is second order as exemplified by $\delta_Z (T)$ in Fig.~\ref{La2Ca2MnO7_muon_parameters_ZF}. 

Recalling that a true long-range order is excluded by the Mermin-Wagner theorem at finite temperature,\cite{Mermin66a,Mermin66b} a mechanism is required to explain the observed transition. Since we have uncovered a spin gap about $80 \, \%$ of the exchange energy, an appreciable breaking of the continuous symmetry in spin space occurs. This is probably the origin of the transition at a non-zero temperature. As noticed by Bao and collaborators, a culprit for the symmetry breaking is the ligand field of the octahedrally-coordinated Mn$^{4+}$ which should lead to a single-ion anisotropy of three-fold axial symmetry perpendicular to the triangular plane.

\section{Conclusions} 
\label{Conclusions}

In summary, the two parameters of a minimal Hamiltonian for the description of La$_2$Ca$_2$MnO$_7$ have been determined. Missing entropy points out to strong paramagnetic short-range correlations. We have shown that the critical paramagnetic spin dynamics is not fully accounted for by spin-wave excitations. The spin dynamics in the order state is anomalously slow with seemingly two frequency scales. In fact, it could be of sporadic nature with the smaller frequency renormalized.  

The short-range correlations and the exotic spin dynamics in the ordered state could be related to the presence of molecular spin substructures.  The missing entropy in the paramagnetic state could have the same origin. 

Clearly, additional $\mu$SR and neutron scattering experiments are worthwhile. The $\mu$SR field response in the ordered state has to be investigated. Additional inelastic neutron work would allow to refine the two Hamiltonian parameters. Neutron scattering measurements have the potentiality to uncover molecular spin substructures.\cite{Lee02}

Since La$_2$Ca$_2$MnO$_7$ is a undistorted triangular antiferromagnet, i.e.\ a two-dimensional spin system, with a relatively 
well determined Hamiltonian, it can be considered as a model system. It should be more amenable to theoretical modeling than 
three-dimensional geometrically frustrated spin systems. 

We thank the European Science Foundation through the Highly Frustrated 
Magnetism program for partial support. The $\mu$SR experiments were performed at the 
Swiss Muon Source, Paul Scherrer Institute, Villigen, Switzerland,

\bibliography{reference.bib}
\bibliographystyle{ws-spin}

\end{multicols}

\end{document}